\documentclass[a4paper,11pt]{article}
\pdfoutput=1
\usepackage{jheppub,hyperref,graphicx,floatflt}
\usepackage{amsmath,latexsym,amssymb,slashed}
\usepackage{enumerate}
\usepackage{mathrsfs}
\usepackage[colorlinks=true,%
            linkcolor=blue,%
            citecolor=blue,%
            urlcolor=blue]{hyperref}

\usepackage{graphicx} % inclusion des figures
\usepackage{amsfonts}
\usepackage{amsmath} % collection de symboles mathématiques
\usepackage{amssymb} % collection de symboles mathématiques
\usepackage{float}

\usepackage{tipa}
\usepackage{textcomp}
\usepackage[T1]{fontenc}

\usepackage{tabularx} % gestion avancée des tableaux

\usepackage{psfrag} % remplacement du texte d'une figure ps par du texte latex

\usepackage{dsfont}
\usepackage{multirow}
\usepackage{color}
\newcommand\beal{\begin{align}}
\newcommand\nn{\nonumber}

\newcommand{\eq}[1]{\begin{equation}#1\end{equation}}
\newcommand{\spl}[1]{\begin{split}#1\end{split}}
\newcommand{\al}[1]{\begin{align}#1\end{align}}

\newcommand{\beq}{\begin{equation}}
\newcommand{\eeq}{\end{equation}}
\def\bea#1\eea{\begin{align}#1\end{align}}
\def\beal#1\eeal{\begin{subequations}\begin{align}#1\end{align}\end{subequations}}
\newcommand{\w}{\wedge}

\def\d{\text{d}}

\def\slashchar#1{\setbox0=\hbox{$#1$}           % set a box for #1
\dimen0=\wd0                                 % and get its size
\setbox1=\hbox{/} \dimen1=\wd1               % get siste of /
\ifdim\dimen0>\dimen1                        % #1 is bigger
\rlap{\hbox to \dimen0{\hfil/\hfil}}      % so center / in box
#1                                        % and print #1
\else                                        % / is bigger
\rlap{\hbox to \dimen1{\hfil$#1$\hfil}}   % so center #1
/                                         % and print /
\fi}

\def\del {\partial}
\def\w {\wedge}

\def\f {{\rm \texttt{f}}}

\usepackage{xcolor}
\usepackage{color}

\newcommand{\cnote}[1]{}

\title{A new mechanism for symmetry breaking from nilmanifolds}

\author[a]{David Andriot,}
\author[b]{Alan Cornell,}
\author[c]{Aldo Deandrea,}
\author[c]{Fabio Dogliotti,}
\author[c]{Dimitrios Tsimpis}

\affiliation[a]{Institute for Theoretical Physics, TU Wien, Wiedner Hauptstrasse 8-10/136, A-1040 Vienna, Austria}
\affiliation[b]{Department of Physics, University of Johannesburg, PO Box 524, Auckland Park 2006, South Africa.}
\affiliation[c]{IP2I, Universit\'e de Lyon, UCBL, UMR 5822, CNRS/IN2P3 4 rue Enrico Fermi, 69622 Villeurbanne Cedex, France}

\emailAdd{david.andriot@tuwien.ac.at}
\emailAdd{acornell@uj.ac.za}
\emailAdd{deandrea@ipnl.in2p3.fr}
\emailAdd{dogliotti@ipnl.in2p3.fr}
\emailAdd{tsimpis@ipnl.in2p3.fr}

\abstract{We present a method to obtain a scalar potential at tree level from a pure gauge theory on nilmanifolds,  a class of negatively-curved compact spaces, and discuss the spontaneous symmetry breaking mechanism induced in the residual Minkowski space after compactification at low energy. We show that the scalar potential is completely determined by the gauge symmetries and the geometry of the compact manifold. In order to allow for simple analytic calculations we consider three extra space dimensions as the minimal example of a nilmanifold, therefore considering a pure Yang-Mills theory in seven dimensions.
}

\begin{document}
\maketitle
\flushbottom
\setcounter{footnote}{0}
\renewcommand{\thefootnote}{\arabic{footnote}}
\setcounter{section}{0}

%%%%%%%%%%%%%%%%%%%%%%
\section{Introduction}
%%%%%%%%%%%%%%%%%%%%%%
\par The number of theories beyond the Standard Model (BSM) has remained incredibly diverse, even in light of the range of experimental results of the past decade. Despite the attractiveness of the idea of a high energy theory allowing to obtain all the fundamental interactions, and the efforts towards reaching such a goal, it is  not yet possible to obtain in a unique and well defined way the Standard Model (SM) of particle physics from fundamental principles alone. Starting instead from the low energy side allows us to implement in the model building the known experimental facts, but typically lacks the uniqueness and the predictivity of a complete fundamental theory.  Somewhere in the middle, and taking inspiration from string theory, compactification of the extra space dimensions can allow us to partially investigate some of the questions which are relegated to free parameters in an effective theory at lower energy.  A familiar example are extra compact dimensions which can be large with respect to the Planck scale, and which may lead to some measurable effects \cite{Antoniadis:1990ew,ArkaniHamed:1998rs,Antoniadis:1998ig,Giudice:1998ck,ArkaniHamed:1999dc}. Compact dimensions used in this setup are usually flat. Also positively-curved compact spaces are used, such as the D-sphere and  discrete quotients thereof (see for example \cite{Cacciapaglia:2016xty}). A second central feature in many of these theories are symmetry breaking mechanisms, like the so called gauge-Higgs models \cite{Manton:1979kb,Hosotani:1983xw,Haba:2004qf}.  This first class of models refers to flat geometry, and generates the scalar sector from the gauge one via quantum fluctuations, giving rise to an effective potential starting at loop level.
Another well-known class of models implementing the same gauge-Higgs idea are those of Randall-Sundrum warped geometries \cite{Randall:1999ee}. This class of gauge-Higgs models has been extensively studied both in the electroweak sector \cite{Hosotani:2005nz,Medina:2007hz} and in the application to grand-unified theories \cite{Hosotani:2008tx,Hosotani:2015hoa}.
Radiative breaking of the electroweak symmetries is also studied in brane models \cite{Antoniadis:2000tq}.

\par In this paper we take a different, but not completely unrelated starting point, and discuss a new method for symmetry breaking which arises from geometry in a new way. Our proposal starts with the hypothesis that there are only gauge bosons, and possibly fermions, in a higher dimensional setup including negatively-curved compact spaces. In order to discuss the idea independently of specific models and fermion representations, we consider here a pure Yang-Mills theory in seven dimensions. Four of them are the usual Minkowski space, the additional three are given by a specific geometry (a {\it nilmanifold}). Here seven is the minimal number of dimensions, related to the fact that we are using  nilmanifolds, but the study can be extended to other cases.

\par Another strong motivation for this set-up and for gauge-Higgs models in general is naturalness. The principle of naturalness expresses the belief that a small parameter can not be an accident, but rather a consequence of symmetry. We do not enter here into a detailed description of this subject, as one can simply note that quantum field theories with gauge fields and fermions are the only natural ones, that is, those containing interacting elementary scalar fields are not. The reason is that the mass of an elementary scalar field is not associated with any approximate symmetry. In the gauge-Higgs framework, the extra-dimensional gauge symmetry protects the smallness of the Higgs boson mass. However, note that the standard gauge-Higgs paradigm also has problems.  Fermions in higher dimensional theories lead to vector-like theories in four dimensions, not to chiral fermions. This can be corrected if the extra dimensions have a non-trivial topology or non-vanishing flux. Moreover the radiatively generated Higgs mass is typically smaller than the measured one. Introducing a warped extra dimensional space typically allows one to have a more satisfactory phenomenology for gauge-Higgs models, but typically at the price of introducing again some fine-tuning of the parameters of the model \cite{Chang:1999nh}.

\par Our setup strongly differs from the one of gauge-Higgs models: as we shall see, in our case the scalar sector (although closely related to the gauge one) stems rather from geometry, and the compactification generates a tree level potential. The gauge symmetry and the geometry fully determine the  properties of the scalar sector of the theory, both in terms of possible representations, couplings and potential.

\par The choice of negatively-curved compact manifolds is suggested by the accumulated results indicating their interesting properties in extra-dimensional models \cite{Kaloper:2000jb}, like the explanation of the hierarchy between the Planck scale and the electroweak scale. For example, compact hyperbolic manifolds (which are special cases of negatively-curved spaces) have two typical length scales: $l_c$ linked to local properties such as the curvature, and $l_G$, related to global properties such as the largest distance on the manifold. The volume grows exponentially with the ratio $l_G/l_c$, which is a topological invariant, allowing for a natural solution to the hierarchy problem  \cite{Orlando:2010kx}. Moreover, compact hyperbolic manifolds of dimension greater than two have the remarkable property of rigidity, implying the absence of all geometric moduli other than the radion (the curvature radius). Hence moduli stabilization reduces to the problem of stabilizing the radion \cite{Nasri:2002rx}.

\par The nilmanifolds considered here, a class of negatively-curved spaces,  are simple and calculable examples thereof, while at the same time being non trivial. Also known as twisted tori, they appeared in the string theory literature as interesting examples of internal manifolds in compactifications to a four-dimensional spacetime \cite{Kachru:2002sk, Grana:2006kf, Caviezel:2008ik, Andriot:2015sia}; they also played a crucial role for T-duality and non-geometric backgrounds (see e.g.~\cite{Andriot:2012vb}). Building on the results of our work  on the three-dimensional Heisenberg nilmanifold \cite{Andriot:2016rdd, Andriot:2018tmb}, we aim at constructing an effective theory describing the scalar and gauge sector. In order to obtain a four dimensional effective theory we need to understand the propagating modes on the nilmanifold. This is done by solving the eigenvalue problem of the Laplacian on this space, using a procedure that produces an infinite series of modes coupled to the gauge sector. These modes are then truncated in order to obtain an effective action at low energy. The spectrum was first studied in the scalar case \cite{Andriot:2016rdd} and later for gauge bosons \cite{Andriot:2018tmb}. The model gives three scalar fields coupled to the gauge, and an interaction potential between the scalars, already at the lowest level of truncation. The potential can be then expanded around a minimum, giving the final action and the masses for the particles.

\par The remainder of this paper is organized as follows: in section \ref{sec2} we first discuss the compactification and truncation in order to obtain the effective four dimensional theory at tree level. We then study the effective model in section \ref{sec3} and discuss different gauge groups and their symmetry breaking patterns. Possible applications to model building are developed in sections \ref{sec4} and \ref{sec5}, and we conclude with some open directions in section \ref{sec6}.

%%%%%%%%%%%%%%%%%%%%%%%%%%%%%%%%
\section{Reduction of 7-dimensional Yang-Mills theory to 4 dimensions}\label{sec2}
%%%%%%%%%%%%%%%%%%%%%%%%%%%%%%%%
\par We start by giving the seven-dimensional (7d) pure Yang-Mills action in the adjoint of  a general Lie group $G$:
\beq
{\cal S} = \int_{M_7} {\rm Tr} ({\cal F}\w *_7 {\cal F}) = \int_{M_7} {\rm Tr} (t_a t_b)\,  {\cal F}^a\w *_7 {\cal F}^b \ ,
\eeq
where $\cal F$ is the field strength, a two-form that takes values in the Lie algebra of $G$, the $t_a$'s are a basis of the Lie algebra under consideration (Einstein's convention on the summation of indices is implied), and $*_7$ is the Hodge star in seven dimensions. The integral is taken over ${M_7}$, a 7d space that will  be specified in the following. The field strength  is expressed in terms of the gauge potential $\cal A$ as:
\beq
{\cal F}= t_a \left(\d {\cal A}^a + \tfrac{1}{2} f^a{}_{bc} {\cal A}^b \w {\cal A}^c  \right) \ , \quad {\cal F}^a_{MN}= 2 \del_{[M} {\cal A}^a_{N]} + f^a{}_{bc} {\cal A}^b_M {\cal A}^c_N  \ ,
\eeq
where the indices $M,N$ run through the different coordinates on the 7d space and $f^a{}_{bc}$ are the structure constants defined as:
\beq
\left[t_b,t_c \right] = t_af^a{}_{bc} ~.
\label{algebra}
\eeq
Let us now turn to the space ${M_7}$ on which the theory is defined.

%%%%%%%%%%%%%%%%%%%%%%%%%
\subsection{Nilmanifolds}
%%%%%%%%%%%%%%%%%%%%%%%%%
\par We will consider the total space ${M_7}$ to be the direct product of two spaces: one is the usual four dimensional Minkowski space, the other is a three dimensional compact space called a nilmanifold. In all generality, a compact nilmanifold is a differentiable manifold that is diffeomorphic to a quotient space of the form $N/\Gamma$, where $N$ is a nilpotent Lie group and $\Gamma$ is a discrete subgroup. More specifically, we will consider the three dimensional nilmanifold built from the Heisenberg algebra defined by:
\al{\label{heisenberg algebra}
[V_1,V_2]=-\f V_3, \quad [V_1,V_3]=[V_2,V_3]=0~,
}
such that $\f\in \mathbb{R}$ is the only non-vanishing  structure constant. This algebra is not to be confused with the Lie algebra associated to the principal bundle $G$ whose structure constants are defined in Eq.~\eqref{algebra}. From this we can construct a one-form basis of the cotangent space that satisfies the Maurer-Cartan equations:
\al{\label{maurerCartan}
\d e^3 = \f e^1 \wedge e^2, \quad \d e^1 = 0, \quad \d e^2 = 0~.
}
Following the conventions of Ref.~\cite{Andriot:2016rdd}, we can parameterize these Maurer-Cartan forms in the following way:
\al{
e^1 = r^1 \d x^1 , \quad e^2 = r^2 \d x^2, \quad e^3 = r^3 (\d x^3 + N x^1\d x^2), \quad N= \frac{r^1 r^2}{r^3}\f \in \mathbb{Z}^* \label{eaf}
~,}
where the $x^i \in [0,1]$ are angular coordinates and the $r^i$ are constant length parameters (radii). We can check that this parametrization is in agreement with Eqs.~\eqref{maurerCartan}. Finally, we use the following discrete identifications:
\al{
x^1 \sim x^1 + n^1, \quad x^2 \sim x^2 + n^2, \quad x^3 \sim n^3 - n^1N x^2, \quad n^1, n^2, n^3 \in\lbrace0,1 \rbrace~.
}
These identifications can be understood as taking the quotient by the action of the discrete subgroup $\Gamma$ in order to make the manifold compact. Using the flat metric $ds^2  = \delta_{ab}e^a e^b$, we see that $\sqrt{g}= r^1 r^2 r^3$ and so the volume is expressed as:
\al{
V=\int d^3x \sqrt{g} =r^1 r^2 r^3~.
}
A word should be said about the geometrical interpretation of such a space. The identifications we made indicate that the space is an $S^1$ fibration over the $T^2$ torus. The trivial fibration of  $S^1$ over $T^2$  is simply $T^3$, but this is not quite the case at hand. Instead, here we have a twisted fibration, the ``twisting'' of the fiber being indicated by the integer $N$. If we take $N=0$, we recover the usual case of $T^3$.

\par Let us go back to our Yang-Mills action. Inserting the expression for the field strength  in terms of the gauge potential, allows us to rewrite the action in the following way:
\bea
{\cal S} &= {\cal S}_2 + {\cal S}_3 + {\cal S}_4 \ , \\
{\rm where}\quad  {\cal S}_2 &= \int_7 {\rm Tr} (t_a t_b)\, \d {\cal A}^a \w *_7 \d {\cal A}^b \ , \\
{\cal S}_3 &=  2 \int_7 {\rm Tr} (t_a t_b) \tfrac{1}{2} f^b{}_{cd}\, \d {\cal A}^a \w *_7 ( {\cal A}^c \w {\cal A}^d ) \ , \\
{\cal S}_4 &= \int_7 {\rm Tr} (t_a t_b) \tfrac{1}{4} f^a{}_{cd}\, f^b{}_{ef} {\cal A}^c \w {\cal A}^d \w *_7 ( {\cal A}^e \w {\cal A}^f ) \ .
\eea
Now we can choose to separate the gauge potential as a sum of 4d-forms and 3d-forms:
\al{
{\cal A}^a &= {\cal A}^a_{M}(x^M) \d x^{M} = {\cal A}^a_{\mu}(x^M) \d x^{\mu} + {\cal A}^a_{m}(x^M) \d x^{m} \\
&= \sum_I U_I(x^m)A^{aI}_{\mu}(x^{\mu}) \d x^{\mu} + \phi^{aI}(x^{\mu}) B_{Im}(x^m) \d x^m \ ,
}
where $M=0,\dots,6$, $\mu=0,\dots,3$ and $m=4,5,6$. In a shorter notation, we can write the gauge potential as:
\beq
\label{decomposition}
{\cal A}^a = \sum_I A^{aI} U_I + \phi^{aI} B_I \ ,
\eeq
where $U_I$ and $ B_I$ are respectively 3d eigenscalars and 3d eigen-one-forms of the Laplacian on the nilmanifold, while $A^{aI}$ and $\phi^{aI}$ are a 4d one-form and a 4d scalar respectively. The $I$ is a multi-index that sums over the basis of 3d eigenforms. The analytical expression for these eigenforms was found in Ref.~\cite{Andriot:2018tmb}. By the property of Laplacian eigenforms, we have:
\beq
*_3 \d *_3 \d U_I = \lambda_{U_I} U_I \ , \ *_3 \d *_3 \d B_I = \lambda_{B_I} B_I \ ,\ \lambda_{U_I} \leq 0 \ , \ \lambda_{B_I} \geq 0 \ ,
\eeq
where the conditions on the signs of the eigenvalues come from the explicit solution of the system, and we have restricted to co-closed one-forms. The development on the eigenform basis is explained in detail in Ref.~\cite{Andriot:2018tmb}. After certain manipulations,\footnote{One needs in particular to decompose the 7d Hodge star into 4d and 3d ones. To that end, the following definitions and properties are useful: we first take $\epsilon_{0123456} = 1 $, and have $(*_3)^2=1,\ *_3 {\rm vol}_3 = 1$, and $(*_4)^2 A_p = (-1)^{p+1},\ *_4 1 = {\rm vol}_4,\ *_4 {\rm vol}_4 = -1$ with $A_p$ a 4d form and $B_p$ a 3d form. We also have $*_7 A_p = *_4 A_p \w {\rm vol}_3$, $*_7 B_p = {\rm vol}_4 \w *_3 B_p $, and $*_7(A_1 \w B_1) = - *_4 A_1 \w *_3 B_1$.} we get the following expressions for the different parts of the action:
\bea
{\cal S}_2  = \int_4 {\rm vol}_4\ {\rm Tr} (t_a t_b)  \bigg(& \big(2 \del_{[\mu} A_{\nu]}^{aI} \del^{\mu} A^{bJ\, \nu} - \lambda_{U_J} A_{\mu}^{aI} A^{bJ\, \mu} \big)\ Y_{U_I U_J}  \\
& + \big( \del_{\mu} \phi^{aI} \del^{\mu} \phi^{aJ} +  \lambda_{B_J} \phi^{aI} \phi^{aJ} \big)\ Y_{B_I B_J} \bigg) \nn\ ,
\eea
\bea
{\cal S}_3  = 2 \int_4 {\rm vol}_4\ {\rm Tr} (t_a t_b) \tfrac{1}{2} f^b{}_{cd} \big(& 2 \del_{[\mu} A_{\nu]}^{aI} A^{cJ\, \mu} A^{dK\, \nu}\ Y_{U_I U_J U_K} \\
& + 2 \del_{\mu} \phi^{aI} A^{cJ\, \mu} \phi^{dK}\ Y_{U_J B_I B_K} \nn \\
& -2 A_{\mu}^{aI} A^{cJ\, \mu} \phi^{dK}\ Y_{U_J dU_I B_K} \nn \\
& + \phi^{aI} \phi^{cJ} \phi^{dK}\ Y_{dB_I B_J B_K} \big) \ , \nn
\eea
\bea
{\cal S}_4  = \int_4 {\rm vol}_4\ {\rm Tr} (t_a t_b) \tfrac{1}{4} f^a{}_{cd} f^b{}_{ef} \big(& 2 A_{[\mu}^{cJ} A_{\nu]}^{dK} A^{eL\, \mu } A^{fM\,\nu}\ Y_{U_J U_K U_L U_M} \\
& + 4 A_{\mu}^{cJ} A^{eL\, \mu} \phi^{dK} \phi^{fM}\ Y_{U_J U_L B_K B_M} \nn\\
& + \phi^{cJ} \phi^{dK} \phi^{eL} \phi^{fM}\ Y_{B_J B_K B_L B_M} \big) \ ,\nn
\eea
where the $Y$'s are given by:
\beq
Y_{U_I U_J} = \int_3 {\rm vol}_3 U_I U_J = \int_3 U_I \w *_3 U_J \ ,\quad Y_{B_I B_J}  = \int_3 B_I \w *_3 B_J \ ;
\eeq
\bea
Y_{U_I U_J U_K} & = \int_3 {\rm vol}_3 U_I U_J U_K \ ,\\
Y_{U_J B_I B_K} & = \int_3  U_J B_I \w *_3 B_K \ ,\\
Y_{U_J dU_I B_K} & = \int_3 U_J \d U_I \w *_3 B_K \ ,\\
Y_{dB_I B_J B_K} & = \int_3 \d B_I \w *_3 (B_J \w B_K ) \ ;
\eea
\bea
Y_{U_J U_K U_L U_M}  & = \int_3 {\rm vol}_3 U_J U_K U_L U_M \ ,\\
Y_{U_J U_L B_K B_M}  & = \int_3  U_J U_L B_K \w *_3 B_M \ ,\\
Y_{B_J B_K B_L B_M}  & = \int_3 B_J \w B_K \w *_3 (B_L \w B_M ) \ .
\eea
A problem remains: the sums over the  indices $I,J,K,L,M$, are infinite sums over the basis of eigenforms. If we want to make this action manageable, a possibility is to organize these terms according to their masses and select only the  light modes.

%%%%%%%%%%%%%%%%%%%%%%%%%%%
\subsection{The truncation}
%%%%%%%%%%%%%%%%%%%%%%%%%%%
\par We would like to organize this infinite series of modes according to their masses, which can be read off from the quadratic terms in ${\cal S}_2$.  To space out the masses as much as possible,  we take the following geometrical limit:
\eq{
\label{limit}
|\f| \ll \frac{1}{r^i}~, ~i=1,2,3 \quad \Rightarrow \quad \frac{1}{r^{1,2}} \ll \frac{1}{r^3}
~.}
This limit, known as the small fiber/large base limit \cite{Andriot:2018tmb}, can be understood by considering the expression for the masses $\lambda_{U_I}$ of the scalar modes, which come in two families:
\bea
\mu^2_{p,q}= p^2\left( \frac{2\pi}{r^1} \right)^2 + q^2\left( \frac{2\pi}{r^2} \right)^2 \\
M^2_{k,l,n}= k^2\left( \frac{2\pi}{r^3} \right)^2 + (2n+1)\mid k \mid \frac{2\pi \mid \f \mid}{r^3}~,
\eea
where  $p,q \in \mathbb{Z}$, $k \in \mathbb{Z}^*$, $n \in \mathbb{N}$, $l=0,1,\dots,\mid k\mid$.
We note that the index $l$ in the second mass is used because of the degeneracy of this eigenvalue. In the case of one-forms, the expression for the masses $\lambda_{B_I}$ come again in two families:
\bea
Y_{\pm}^{p,q} = P^2 + Q^2 + \frac{\f^2}{2} \pm \sqrt{\left(P^2 + Q^2 + \frac{\f^2}{2}\right)^2 -  (P^2 + Q^2)^2} \\
Y_{\pm}^{k,l,n} = M^2_{k,l,n} + 2\mid k\frac{2\pi \f}{r^3} \mid + \frac{1}{2}\f^2 \pm \sqrt{\left( \mid k\frac{2\pi \f}{r^3} \mid + \frac{1}{2}\f^2 \right)^2 + 2(n+1)\mid k\frac{2\pi \f}{r^3} \mid \f^2}~,
\eea
where $P=p2\pi /r^1$, $Q=q2\pi /r^2$. As shown in \cite{Andriot:2018tmb}, we then have the following lightest modes (in a basis of real orthonormal eigenmodes):\\

\noindent{\it Scalars}:
\eq{\spl{
U_{I=1}&=\frac{1}{\sqrt{V}}~;~~~\lambda_{U_1}=0
~,}}
where  we recall that $V=r^1 r^2 r^3$ is the volume of the nilmanifold.\\

\noindent{\it Co-closed one-forms}:
\eq{\spl{
B_{I=1}&=\frac{1}{\sqrt{V}}e^1~;~~~\lambda_{B_1}=0\\
B_{I=2}&=\frac{1}{\sqrt{V}}e^2~;~~~\lambda_{B_2}=0\\
B_{I=3}&=\frac{1}{\sqrt{V}}e^3~;~~~\lambda_{B_3}=\f^2
~,}}
where the $e^a$ one-forms  satisfy the Maurer Cartan equation \eqref{maurerCartan}. We see that the eigenvalues define the masses for the selected modes. Almost all of these are massless except for the  one-form $B_3$, which has a mass $\f^2$ directly related to the geometry, {cf}.~Eqs.~\eqref{heisenberg algebra}, \eqref{eaf}. The decomposition in Eq.~\eqref{decomposition} of the gauge potential simplifies to:
\beq
{\cal A}^a = A^{a} U_1 + \sum_{I=1}^3 \phi^{aI} B_I \ .
\eeq
The resulting non-vanishing couplings are as follows,
\eq{ Y_{U_1 U_1}  =  1 ~,~~~
Y_{B_I B_J}=\delta_{IJ}
~,}
for the quadratic couplings, together with
\bea
Y_{U_I U_J U_K}:\quad & Y_{U_1 U_1 U_1}  =  \frac{1}{\sqrt{V}} \nn\\
Y_{U_J B_I B_K}:\quad & Y_{U_1 B_L B_L} =  \frac{1}{\sqrt{V}} \ ,\ L=1, 2, 3\nn\\
Y_{dB_I B_J B_K}:\quad & Y_{dB_3 B_2 B_1} =-Y_{dB_3 B_1 B_2} =  \frac{\f}{\sqrt{V}}\nn\\
Y_{U_I U_J U_K U_L}:\quad & Y_{U_1 U_1 U_1 U_1}=  \frac{1}{V}\nn\\
Y_{U_I U_J B_K B_L}:\quad & Y_{U_1 U_1 B_L B_L}=  \frac{1}{V}\ ,\ L=1, 2, 3\nn\\
Y_{B_I B_J B_K B_L}:\quad & Y_{B_1 B_2 B_1 B_2}= Y_{B_1 B_3 B_1 B_3}= Y_{B_2 B_3 B_2 B_3}=  \frac{1}{V}\ ,
\eea
where in the last line we can also have anti-symmetric permutations of the first two and/or the last two indices. After the truncation to these light modes, we can finally write the action as:
\eq{S=
\int\d x^4 \text{Tr}\Big(
\tfrac{1}{2} F_{\mu\nu}F^{\mu\nu}+\sum_{I=1}^3D_\mu\phi^{I} D^\mu\phi^{I}+M^2(\phi^{3})^2+\mathcal{U}
\Big)
~,}
where
\eq{
\hspace{-0.2in} \mathcal{U}=\text{Tr}\Big( -2 gM[\phi^1,\phi^2]\phi^3 + \tfrac12 g^2\sum_{I,J=1}^3[\phi^I,\phi^J][\phi^I,\phi^J]\Big)
~,}
with $F_{\mu \nu} = 2 \partial_{ [ \mu} A_{\nu ]} + g [A_{\mu}, A_{\nu}]$ and $D_{\mu} = \partial_{\mu} + g [A_{\mu}, \cdot]$. We have relabelled the parameters such that $g=1/\sqrt{V}$  and $M= \mid \f \mid$.

\par Lastly, we want to perform a set of transformations in order to retrieve the usual Yang-Mills conventions:
\eq{\spl{
t_a &\rightarrow i t_a \\
\eta_{\mu \nu} &\rightarrow -\eta_{\mu \nu}~,
}}
resulting in the final expression:
\eq{S=
\int\d x^4 \text{Tr}\Big(
-\tfrac{1}{2} F_{\mu\nu}F^{\mu\nu}+\sum_{I=1}^3D_\mu\phi^{I} D^\mu\phi^{I}-M^2(\phi^{3})^2-\mathcal{U}
\Big) ~,}
where:
\eq{
 \mathcal{U}=\text{Tr}\Big( -2i gM[\phi^1,\phi^2]\phi^3 + \tfrac12 g^2\sum_{I,J=1}^3[\phi^I,\phi^J][\phi^I,\phi^J]\Big)
~.}
We see that the original seven dimensional pure Yang-Mills theory gives rise to, upon compactification and truncation to the light modes, a low energy  effective action consisting of a four-dimensional Yang-Mills coupled to three scalar fields in the adjoint representation. Moreover, one of these scalars is massive and all three scalars interact via the potential. The next step is to understand the structure of the potential in order the find a vacuum of the theory, i.e.~a local minimum of the potential. This is generally-speaking not a trivial question. The potential has a total number of independent variables that are three times the dimension of the Lie algebra associated to $G$, meaning that even for low dimensional Lie algebras, the number of real independent variables can be quite large.

%%%%%%%%%%%%%%%%
\section{The vacuum structure}\label{sec3}
%%%%%%%%%%%%%%%%
\par To obtain the masses of the fields we must diagonalize the mass matrix of the quadratic fluctuations  around the vacuum configuration. First, let us define the potential we want to study as the sum of the interaction terms and the mass term. Explicitly:
\eq{
\frac{\mathcal{V}}{M^2}=\text{Tr} (\phi^{3})^2  +\frac{\mathcal{U}}{M^2}~.	
}
A vacuum configuration is a solution of:
\begin{eqnarray}
\frac{\mathcal{V}(\phi +\delta \phi)-\mathcal{V}(\phi)}{M^2} &=& \text{Tr}\left( 2\phi^3\delta \phi^3 \right) - 2i\frac{g}{M} \text{Tr} \Big([\phi^1,\phi^2]\delta \phi^3 + [\phi^3,\phi^1]\delta \phi^2 + [\phi^2,\phi^3]\delta \phi^1 \Big) \nonumber \\
&&\quad \quad +2 \frac{g^2}{M^2} \text{Tr} \Big( \sum_{I,J=1}^3[\phi^I,\phi^J][\phi^I,\delta \phi^J]\Big)
=0 \ .
\end{eqnarray}
We can see the theory contains a class of vacua of the form $\phi^{I}=\text{constant}$, where,
\eq{\label{gclass}
\phi^{3}=0~;~~~
[\phi^{1},\phi^{2} ]=0~.
}
Once a vacuum  $\phi^I=\phi^I_0$ is found, infinite classes of other vacua are generated by conjugation of arbitrary elements $U \in G$, $\phi^I_0\rightarrow U^{\dagger}\phi^I_0 U$. Since this conjugation is a symmetry of the potential, it is a map from vacua to vacua. Moreover, if the vacuum $\phi^I_0$ is not invariant under the conjugation, it will be mapped to a different vacuum with the same energy. This is of course the Goldstone mechanism, implying the existence of various massless 4d scalars in the vacuum, as we will confirm explicitly in the following.

\par Now that a class of minima has been identified in \eqref{gclass}, we can   develop the potential to second order in the variations of the fields. The variation of the potential reads:
\begin{eqnarray}
\frac{\delta^2\mathcal{V}}{2M^2}&=&  \text{Tr} ( \delta \phi^3)^2 - \frac{ig}{M} \text{Tr} \left( \left[\delta \phi^1, \phi_{0}^2 \right]\delta \phi^3 + \left[ \phi_{0}^1, \delta \phi^{2} \right]\delta \phi^3 \right) \label{del2V} \\
&& + \frac{g^2}{M^2} \text{Tr} \left( \left[\delta \phi^1,\phi_{0}^2 \right]^2 + \left[\delta \phi^2,\phi_{0}^1 \right]^2 + \left[\delta \phi^3,\phi_{0}^1 \right]^2 + \left[\delta \phi^3,\phi_{0}^2 \right]^2 + 2\left[\delta \phi^1,\phi_{0}^2 \right] \left[\phi_{0}^1,\delta \phi^2 \right]\right)
~, \nonumber
\end{eqnarray}
where $\phi_{0}^1$ and $\phi_{0}^2$ are the values of the fields at the minimum. This expression gives a matrix of dimension $(3\times\text{dim}(Lie (G))^2$, where dim$(Lie(G))$ is the dimension of the Lie algebra associated to the gauge group $G$. In order to obtain the masses of the fields around this minimum we have to diagonalize this matrix once a specific gauge group has been chosen. We will now discuss the case $G=SU(3)$.

%%%%%%%%%%%%%%%%%%%%%%%%%%%%%%%%%%%%%%%%%%%%
\section{From $SU(3)$ to $SU(2)\times U(1)$}\label{sec4}
%%%%%%%%%%%%%%%%%%%%%%%%%%%%%%%%%%%%%%%%%%%%
\par The first case of interest is $SU(3)$:   it is a minimal setup that contains $SU(2) \times U(1)$, which is suitable to model the electroweak sector. Also, its Lie algebra is eight dimensional, making it rather manageable. We denote by $\mathfrak{su}(3)$ the Lie algebra associated to $SU(3)$. A matrix $A \in \mathfrak{su}(3)$ is characterized, in our conventions, by $A=A^{\dag}$ and Tr$(A)=0$. We shall use the Gell-Mann convention for the  $SU(3)$ generators  \cite{GellMann:1962xb}.  As already remarked, the vacuum solution that we choose may have some residual symmetries. These symmetries will determine the residual gauge symmetry after the vacuum solution has been selected.

\par Let us start by explaining how to choose the vacuum in order to have an unbroken  $SU(2) \times U(1)$ gauge: we want that $\phi_0^1$ and $\phi_0^2$ commute, in order for them to satisfy condition \eqref{gclass} for a minimum. We can thus simply choose them as combinations of the diagonal Gell-Mann matrices, which we know commute: indeed the diagonal Gell-Mann matrices form the Cartan subalgebra of $\mathfrak{su}(3)$,  i.e.~its maximal Abelian subalgebra. A simple way to parameterize our vacuum is to take $\phi_0^1=\phi_0^2$ and take $\phi_0$ to be a generic element of the Cartan subalgebra:
\al{
\phi_{0}^1=\phi_{0}^2=\phi_0= \begin{pmatrix}
a & 0 & 0 \\
0 & b & 0 \\
0 & 0 & c
\end{pmatrix}
~,}
with $a,b,c$ real parameters and $a+b+c=0$, since $\phi_0 \in \mathfrak{su}(3)$. We want to leave an $\mathfrak{su}(2)$ subalgebra unbroken, in other words, we want that $\phi_0$ commutes with the elements that generate an $\mathfrak{su}(2)$. If we take $a=b$, so that the trace condition imposes $c=-2a$, it follows  that $\phi_0\propto \lambda_8$.
Only one parameter remains free, the normalization of the field.
In this case $\phi_0$
 commutes with the Gell-Mann matrices
$\lambda_1$, $\lambda_2$, $\lambda_3$ and $\lambda_8$, which generate an  $\mathfrak{su}(2)\oplus\mathfrak{u}(1)$ subalgebra of $\mathfrak{su}(3)$.
\par We need to diagonalize the matrix \eqref{del2V} of second derivatives of the potential at the point $\phi^3=0, \phi^1=\phi^2 = \phi_0$ (this is a $24\times24$ matrix). The result is a list of masses, where the  multiplicity of each mass indicates the dimension of the representation of the new gauge group (i.e.~after symmetry breaking) in which this mass transforms. In our case we obtain the following masses for the scalar fields:
\al{
&\text{12 massless degrees of freedom (dof)}\\
&M_0^2=2M^2 \quad \text{(4 dof)} \\
&M_{\pm}^2=M^2 \left( 1 + 144 a^2 \frac{g^2}{M^2} \pm \sqrt{1 + 288 a^2 \frac{g^2}{M^2}} \right) \quad \text{(4+4 dof)}
~,}
where in the last line (4+4 dof) means that the subspace associated to the mass $M_+^2$ is four dimensional, and likewise for the one associated to $M_-^2$.

\begin{figure}
\label{graphmasses}
\centering
\includegraphics[scale=.75]{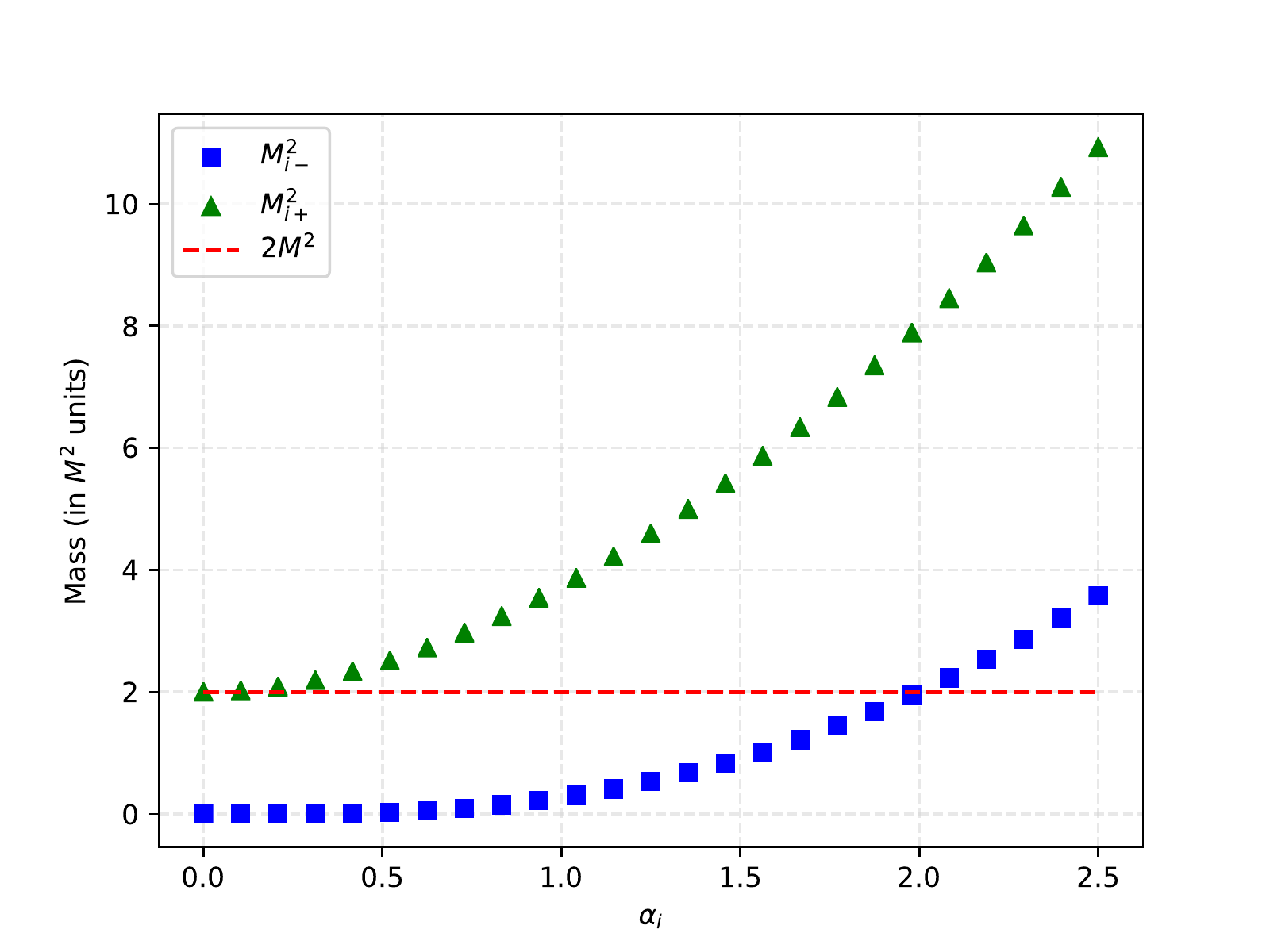}
\caption{Graph of the three types of scalar masses of the $SU(3)$ model. This graph is relevant for the other models too since all the masses are of the form $M_{i\pm}^2 = 1 + \alpha_i^2 \pm \sqrt{1 + 2\alpha_i^2}$ with $\alpha_i$ being linear combinations of the vacuum parameters. We can see that $M_{i-}^2$ goes to zero for small $\alpha_i$ while the two other masses remain around the $M^2$ scale.}
\end{figure}

\par Let us see how to associate each mass to a representation. From the eigenvectors (not listed here) one can read that the masses $M_{\pm}^2$ are in a (2,1) representation of the gauge group $SU(2)\times U(1)$. The mass $M_0^2$ is in the adjoint representation of $SU(2)\times U(1)$. Lastly, we can deduce the representation of the massless modes by a counting argument. We started with  three copies of the adjoint representation of $SU(3)$. We used twice the two- (complex) dimensional representation for the masses $M_{\pm}^2$. We used one adjoint representation for $M_0^2$. We are therefore left with two copies of the adjoint representation and one of the two-dimensional representation. The total dimension of these representations is $2\times 4+4=12$, just like the multiplicity of the massless modes.

\par There is also a more intuitive reasoning that could have let us guess the solution: it can be seen that the masses $M_{0}^2=2M^2$ are in the adjoint representation, since we chose $\phi_0$ to commute with the generators of $\mathfrak{su}(2)$ and $\lambda_8$ that generates $U(1)$. We would therefore expect the mass of these states not to be modified by the vacuum we selected. Hence the $\mathfrak{su}(2)\oplus \mathfrak{u}(1)$ part of $\phi^3$ remains with a mass $M^2$ (the factor 2 being simply a consequence of the square). The other two massive states depend on the parameter $a$ and are in the only representation that doesn't commute with $\phi_0$, namely the two-dimensional one. We will see that this pattern persists for bigger gauge groups.

Finally, we can also calculate the mass of the gauge bosons corresponding to the broken generators. It is simply given by the coupling terms between the scalars and the gauge fields:
\al{
g^2\text{Tr}\left([A_{\mu},\phi_0]^2\right)=-18 g^2a^2 \sum_i  ( A^i_{\mu})^2~,
}
where on the right-hand side the sum is over the gauge fields associated to the broken generators $\lambda_4$, $\lambda_5$, $\lambda_6$, $\lambda_7$.

Now that all the masses have been determined, the effective theory is fully determined. However there is still some freedom in the hierarchy of the masses. Let us now turn to this point: the model has two scalar fields in a (2,1) representation of the gauge, that correspond to the masses $M_-^2$ and $M_+^2$. These representations are SU(2) doublets, suitable for the description of the Higgs field. The mass $M_-^2$ is a difference of two terms. In the limit where $a^2g^2/M^2 \rightarrow 0$, this mass also goes to zero, while $M_+^2$ will be of order $M^2$. If we consider the region $M^2\gg a^2g^2\gg m_{limit}^2$ where $m_{limit}$ is the maximum detectable mass, we can adjust the ratio $a^2g^2/M^2$ so that $M_-^2 = M_{Higgs}^2$, while all the other fields can have masses well beyond detection.

\par The simple model we have presented allows us to find one scalar field coupled to the electroweak sector whose mass can be taken to be arbitrarily lower than the other masses, therefore allowing us to accommodate the Higgs boson at its measured mass value. This toy model is, however, not the complete gauge sector of the SM, only the electroweak sector. Another important point is the presence of the massless scalar fields. Three massless fields remain in the spectrum, two in the adjoint representation and one in a two-dimensional representation. These massless scalars are coupled to the gauge the same way the Higgs is. Moreover in the SM, the electroweak symmetry breaking is $SU(2)\times U(1)$ spontaneously broken to $U(1)_{em}$. If we want to break the gauge sector along this pattern we can break $SU(2)$ to $U(1)$, but we obtain two copies of $U(1)$, one from the breaking of $SU(3)$, and one from $SU(2)$. This situation is exactly the same as the one encountered in gauge-Higgs models when considering the gauge part in isolation \cite{Antoniadis:2001cv}. Indeed we cannot further break $U(1)$ symmetries in this simple model without introducing extra ingredients and in particular fermions and their interactions. This can be traced back to the  vacuum condition: Eq.~\eqref{gclass}  requires  $\phi^1_0$ and $\phi^2_0$ to commute, which can therefore be taken to be in the Cartan subalgebra of $SU(3)$. On the other hand, the part of the gauge that is broken is the one that does not commute with the vacuum solution. As a consequence the Cartan subalgebra (two copies of $U(1)$ in the case at hand) always remains unbroken.

%%%%%%%%%%%%%%%%%%%%%%%%
\section{Other examples}\label{sec5}
%%%%%%%%%%%%%%%%%%%%%%%%
\par The breaking pattern we have discussed can be used in practice for model building, for example in the breaking of unification groups, but requires some extra ingredients to take care of the residual symmetries and zero modes. In the following we will discuss examples of  symmetry groups in the adjoint representations (as we have a Yang-Mills theory) which contain $\mathfrak{su}(2)$ doublets as subalgebras. This is in order  to allow for  the possibility to obtain a Higgs-like potential.   However, if not restricted by this requirement, many other breaking patterns can be studied.

%%%%%%%%%%%%%%%%%%%%%%%%%%%%%%%%
\subsection{The case of $SU(5)$}
%%%%%%%%%%%%%%%%%%%%%%%%%%%%%%%%
\par In this section we study an $SU(5)$ gauge symmetry that breaks spontaneously to  $SU(3)\times SU(2) \times U(1)$. We start again by giving the explicit matrix representation of $\mathfrak{su}(5)$, where the general element of $\mathfrak{su}(5)$ is of the form:
\al{
\begin{pmatrix}
a_1 +c & a_4 & a_5   & h_1 & h_2 \\
a_4^* & a_2 + c & a_6 & h_3 & h_4 \\
a_5^* & a_6^* & -(a_1+a_2) + c & h_5 & h_6 \\
h_1^* & h_3^* & h_5^* & b_1 -\frac{3}{2}c & b_3 \\
h_2^* & h_4^* & h_6^* & b_3^* &  -b_1 -\frac{3}{2}c
\end{pmatrix}~.
\label{su(5)}
}
This matrix is traceless and Hermitian (the diagonal elements are taken to be real) and therefore lies in $\mathfrak{su}(5)$. The parameters have been chosen so that the decomposition is made explicit. Indeed, the $a$'s generate an $\mathfrak{su}(3)$ algebra: the $b$'s generate an $\mathfrak{su}(2)$, the $c$ a $\mathfrak{u}(1)$, while the $h$'s are in the (3,2)- (complex) dimensional representation of respectively $\mathfrak{su}(3)$ and $\mathfrak{su}(2)$. The $h$'s are also charged under $\mathfrak{u}(1)$. We see that this situation is very similar to the one we had before for $SU(3)$. Again, we want to find a vacuum $\phi^1=\phi^2=\phi_0$ such that this vacuum commutes with the generators of the unbroken gauge group. We can construct $\phi_0$ in a similar fashion as before. Let us  take $\phi_0$ to be:
\al{
\phi_0= \begin{pmatrix}
a &   &   &   &  \\
  & a &   &   &  \\
  &   & a &   &  \\
  &   &   & -\frac{3}{2}a &  \\
  &   &   &   & -\frac{3}{2}a
\end{pmatrix}~.
}
The invariance of $\phi_0$ can then be seen at the level of the algebra since:
\eq{
\forall ~\lambda = \begin{pmatrix}
A &  \\
  & B
\end{pmatrix} \in \mathfrak{su}(5)~, ~~  \left[\lambda ,  \phi_0 \right] = 0
~,}
where $A \in \mathfrak{su}(3)$, $B \in \mathfrak{su}(2)$ and of course $Tr(\lambda )= Tr(A)=Tr(B)=0$. Note that $\phi_0$ also commutes with all the generators of the Cartan subalgebra since $\phi_0$ itself is in the Cartan subalgebra.

\par Applying the same procedure as  before  to the now $72\times 72$ mass matrix, we find the masses to be:
\al{
&\text{Massless} \quad \text{(36 dof)} \\
&M_0^2 = 2M^2 \quad \text{(12 dof)} \\
&M_{\pm}^2= M^2 \left( 1+100 a^2\frac{g^2}{M^2} \pm \sqrt{1+200 a^2 \frac{g^2}{M^2}} \right) \quad \text{(12+12 dof)}~.
}
These masses are very similar to the masses found for the $SU(3)$ gauge. The masses $M_+^2$ and $M_-^2$ fall again into the only representation that does not commute with the gauge, namely the (3,2,1) representation generated by the $h$'s in Eq.~\eqref{su(5)}. The mass $M_0^2$ is again in the  adjoint representation of the new gauge. The dimension of this adjoint representation is the sum of the dimensions of each adjoint representation,  $8+3+1=12$. Again, we can compute the mass of the broken gauge bosons that we find to be:
\al{
M_{boson}=\frac{25}{4}g^2a^2~.
}
These also transform in the (3,2,1) representation of the gauge group.

\par The argument about the masses discussed for SU(3) is also valid here, and we can have the field with the mass $M_-^2$ at a much lower scale than the rest of the masses. This
state is also charged under the strong interactions in this case.

%%%%%%%%%%%%%%%%%%%%%%%%%%%%%%%%
\subsection{The case of $SU(6)$}
%%%%%%%%%%%%%%%%%%%%%%%%%%%%%%%%
\par $SU(6)$ has the advantage of breaking into $SU(3) \times SU(2) \times U(1) \times U(1)$, which is close to the SM gauge, and allows for  a (1,2,1,1) representation of the new gauge group (i.e.~after symmetry breaking) at an arbitrary low mass. Quite similarly to the $\mathfrak{su}(5)$ case, we can write a general element of $\mathfrak{su}(6)$ as:
\al{
\begin{pmatrix}
a_1 +c + d & a_4 & a_5   & h_1 & h_2 & l_1\\
a_4^* & a_2 + c +d & a_6 & h_3 & h_4 & l_2\\
a_5^* & a_6^* & -(a_1+a_2) + c+d & h_5 & h_6 & l_3\\
h_1^* & h_3^* & h_5^* & b_1 -\frac{3}{2}c+d & b_3 & p_1\\
h_2^* & h_4^* & h_6^* & b_3^* &  -b_1 -\frac{3}{2}c+d & p_2\\
l_1^* & l_2^* & l_3^* & p_1^* & p_2^* & -4d
\end{pmatrix}.
\label{su(6)}
}
Again the parameterization makes explicit the decomposition. The $a$'s generate an $\mathfrak{su}(3)$, the $b$'s an $\mathfrak{su}(2)$, $c$ and $d$ are two copies of $\mathfrak{u}(1)$, the $h$'s are in the (3,2,1,0), the $l$'s in the  (3,1,2,1), the $p$'s in the (1,2,1,1). For each representation the charge under the two $U(1)$'s is calculated simply by using the commutation relations between the generators of the $U(1)$'s and the generators of the representation. This example is interesting because it is the first example that gives different massive representations of the gauge group. We can parameterize the vacuum in the following way:
\al{
\phi_0= \begin{pmatrix}
a &   &   &     &    &  \\
  & a &   &     &    &  \\
  &   & a &     &    &  \\
  &   &   & b-a &    &  \\
  &   &   &     & b-a&  \\
  &   &   &     &    &-(2b+a)
\end{pmatrix}~.
}
This is the first example where the parameterization of $\phi_0$ leaves two parameters free. The reason is simply that here we decompose $\mathfrak{su}(6)$ into sufficiently many subalgebras, so that even with the trace condition we still have two free parameters. For this vacuum, the diagonalization of the now $105\times 105$ mass matrix gives:
\al{
& \text{Massless} \quad \text{(48 dof)}\\
& M_0^2=2M^2 \text{  (13 dof, rep:(8,3,1,1))} \\
& M_{i\pm}^2 = M^2 \left( 1 + \frac{g^2}{M^2}\alpha_i^2  \pm \sqrt{1+2\frac{g^2}{M^2}\alpha_i^2} \right) \ , \\
\text{where} \nonumber \\
&\alpha_1^2 = 144b^2  \quad \text{(rep:(1,2,1,1))} \\
&\alpha_2^2 = 64(a+b)^2  \quad \text{(rep:(3,1,2,1))}\\
&\alpha_3^2 = 16(2a-b)^2 \quad \text{(rep:(3,2,1,0))}
~.}
The representations can be associated to the masses simply by using the dimension of each representation and the multiplicity of the masses in the diagonalized mass matrix. We can reach the same conclusion by using the following reasoning. Let us first see why $\alpha_1$ is associated with the (1,2,1,1) representation. If we take the limit $b \rightarrow 0$, $\phi_0$ becomes invariant under the group $SU(3)\times SU(3)\times U(1)$. The mass $M_1^2$ goes to zero, and the masses $M_2^2$ and $M_3^2$ become equal. In this case only one non-adjoint representation remains and is a (3,3,1) representation of the gauge $SU(3)\times SU(3)\times U(1)$, generated by the $h$'s and $l$'s in the decomposition \eqref{su(6)}. So we see that the $p$'s generate the mass $M_1^2$. We can use  a similar reasoning in order to associate the correct representations to the masses $M_2^2$ and $M_3^2$. We take the limit where $b \rightarrow 2a$, so that the mass $M_3^2$ goes to zero. The masses $M_1^2$ and $M_2^2$ become equal. The gauge group in this case is $SU(5) \times U(1)$, and of course the massive states are generated by the $l$'s and $p$'s. We already know that the $p$'s were associated to the mass $M_1^2$, therefore we know that the $l$'s are associated to the mass $M_2^2$. Once again, we can calculate the masses of the broken gauge bosons:
\al{
& M_{boson1}^2 =  9g^2 b^2  \quad \text{(rep:(1,2,1,1))},\\
& M_{boson3}^2 = 4g^2 (a+b)^2 \quad \text{(rep:(3,1,2,1))}, \\
& M_{boson2}^2 = g^2 (b-2a)^2  \quad \text{(rep:(3,2,1,0))}~.
}
In order to have the (1,2,1,1) representation at a much lower mass than the other massive representations, all we have to do is take $M^2,a^2 \gg b^2$. Indeed, taking $M^2 \gg b^2$ allows for $M_{1-}^2$ to be very small, while $a^2 \gg b^2$ forces the other masses to be much higher. Just like before, we can then adjust the ratio $9g^2b^2/M^2$ in order to match $M_1^2$ with the mass of the Higgs boson.

%%%%%%%%%%%%%%%%%%%%%
\section{Conclusions}\label{sec6}
%%%%%%%%%%%%%%%%%%%%%
\par We have presented a simple model of the Yang-Mills and Higgs sectors, obtained from the compactification of a pure Yang-Mills theory on a spacetime with extra dimensions of negatively-curved geometry. We showed how the scalar potential is determined by the gauge symmetry and by the geometry of the compact manifold. Contrary to previous gauge-Higgs models, a non vanishing potential is already generated  at tree level, a feature that should lead to improved phenomenological properties.

\par A spontaneous symmetry breaking mechanism is induced  in the residual Minkowski space after compactification at low energy. Our mechanism can be used in Grand Unification models and for electroweak symmetry breaking. The effective theory contains a scalar field whose mass is not fixed by the characteristic mass set by the geometry, and is therefore adjustable. Moreover, the gauge group can be chosen so that  this field is in the same representation as the Higgs field.

\par For concreteness we have taken the extra dimensions to be a three-dimensional nilmanifold, although the model can be extended to different types of manifolds and/or dimensions. In particular there are various different possibilities for the geometry of the extra dimensions within the class of solvmanifolds, of which nilmanifolds are a special case (see \cite{Andriot:2010ju} for a review), that can be explored systematically with tools similar to the ones presented here. Another possible extension concerns the choice of the gauge group: for simplicity we have limited ourselves to $SU(N)$ groups, but this is by no means exhaustive.

\par The model of the present paper is not realistic, as it is lacking the fermion sector altogether. Crucially, incorporating the fermions is expected  to change some of the quantitative properties of the vacuum structure (in particular concerning the flat directions), and will allow to study the anomalies. In particular it should also provide at loop level a mechanism for generating masses for the various moduli (massless scalars), which as we saw are ubiquitous in the examples presented here. These points can be discussed with the same techniques used to address them in the standard gauge-Higgs models. We are planning to report on this in future work.

%\begin{appendix}
%\end{appendix}
%%%%%%%%%%%%%%%%%%%%%%%%%%%%%%%%%%%%%%%%%%%%
\section*{Acknowledgements}
%%%%%%%%%%%%%%%%%%%%%%%%%%%%%%%%%%%%%%%%%%%%
D.A.~acknowledges support from the Austrian Science Fund (FWF): project number M2247-N27. A.S.C.~is supported in part by the National Research Foundation of South Africa (NRF). A.S.C.~thanks the University of Lyon 1 and IP2I for support during the collaboration visit in Lyon.

\providecommand{\href}[2]{#2}\begingroup\raggedright\endgroup

\end{document}